\documentclass[conference,letterpaper]{IEEEtran}
\usepackage[utf8]{inputenc} 
\usepackage[T1]{fontenc}
\usepackage{url}
\usepackage[cmex10]{amsmath}
\usepackage{amsmath}
\usepackage{amsmath,amssymb,amsfonts}
\usepackage{algorithmicx}
\usepackage{algorithm,algpseudocode}
\usepackage{algcompatible}
\usepackage{array}
\usepackage[caption=false,font=normalsize,labelfont=sf,textfont=sf]{subfig}
\usepackage{textcomp}
\usepackage{stfloats}
\usepackage{url}
\usepackage{verbatim}
\usepackage{graphicx}
\usepackage{caption}
\usepackage{cite}
\usepackage{bm}
\usepackage{xcolor}
\usepackage{colortbl}
\usepackage{multirow}
\usepackage{bbm}
\usepackage[figuresright]{rotating}
\allowdisplaybreaks
\newtheorem{example}{Example}
\newtheorem{definition}{Definition}

\ifodd 1

\else

\fi

\begin{document}
\title{Uncoded Download in Lagrange-Coded Elastic Computing with Straggler Tolerance} 

\author{
\IEEEauthorblockN{Xi Zhong\textsuperscript{1},  Samuel Lu\textsuperscript{2},  Jörg Kliewer\textsuperscript{3} and Mingyue Ji\textsuperscript{1}}

\IEEEauthorblockA{\textit{\textsuperscript{1}Department of Electrical and Computer Engineering}, \textit{University of Florida}, Gainesville, FL, USA\\
Email: \{xi.zhong, mingyueji\}@ufl.edu }

\IEEEauthorblockA{\textit{\textsuperscript{2}Rowland Hall St. Marks High School}, Salt Lake City, UT, USA \\
Email: samuellu@rowlandhall.org}

\IEEEauthorblockA{\textit{\textsuperscript{3}Department of Electrical and Computer Engineering}, \textit{New Jersey Institute of Technology}, Newark, NJ, USA\\
Email: jkliewer@njit.edu }}

\maketitle

\begin{abstract}
Coded elastic computing, introduced by Yang \emph{et al.} in 2018, is a technique designed to mitigate the impact of elasticity in cloud computing systems, where machines can be preempted or be added during computing rounds. 
This approach utilizes maximum distance separable (MDS) coding for both storage and download in matrix-matrix multiplications. The proposed scheme is unable to tolerate stragglers and has high encoding complexity and upload cost. In 2023, we addressed these limitations by employing uncoded storage and Lagrange-coded download. However, it results in a large storage size. To address the challenges of storage size and upload cost, in this paper, we focus on Lagrange-coded elastic computing based on uncoded download. We propose a new class of elastic computing schemes, using Lagrange-coded storage with uncoded download (LCSUD). Our proposed schemes address both elasticity and straggler challenges while achieving lower storage size, reduced encoding complexity, and upload cost compared to existing methods. 
\end{abstract}

\section{introduction}
Coded elastic computing is an emerging paradigm designed to address the elasticity of virtual machines in distributed cloud systems, where machines can be preempted or added during computing rounds.
The first coded elastic computing scheme was proposed in \cite{yang2018coded}, using a maximum distance separable (MDS)-coded storage and uncoded download strategy for homogeneous systems, where machines have the same computation speed and storage capacity.
Under this framework, \cite{KSA2021} proposed hierarchical computation assignments aimed at improving speed and straggler tolerance by assigning fewer machines to the initial tasks in the task list, while later tasks are distributed among a greater number of machines.
To address heterogeneous systems, where machines have varying computing speeds and storage constraints, \cite{wcj2021hs} introduced a combinatorial optimization approach to derive an optimal computation assignment, which was extended in \cite{myjpractice} to handle scenarios that combine elasticity and straggler tolerance.
In \cite{DHFL2023}, transition waste was introduced to measure unnecessary changes in task allocation due to elasticity. To mitigate this, the authors proposed shifted cyclic computation assignments, which achieve zero transition waste as long as the number of available machines fluctuates within a predefined range.
 
One of the challenges for the framework in \cite{yang2018coded} is the limitation of certain types of computations (e.g., linear) due to its reliance on linear coding. To address this limitation, some works employed uncoded storage in elastic computing by placing datasets on machines without using coding techniques. 
The authors in \cite{usutec2022} formulated a combinatorial optimization problem and derived optimal solutions to minimize overall computation time for a given uncoded storage.
In \cite{XJM2023}, Lagrange-coded download was used during the download phase for homogeneous systems with uncoded storage. Later, the authors in \cite{XJM2024} extended this uncoded storage and coded download approach to heterogeneous systems, proposing a hierarchical storage placement algorithm to minimize expected computation time and reduce storage requirements. A decentralized uncoded storage elastic computing scheme for heterogeneous systems was proposed in \cite{HYWQJ2024}. This method provides a closed-form solution to achieve optimal computation time without requiring coordination among machines’ storage.

Some existing studies on coded elastic computing, including \cite{yang2018coded, KSA2021, wcj2021hs, myjpractice, DHFL2023, usutec2022}, originally focus on matrix-vector multiplications. For matrix-matrix multiplications, \cite{yangCEC} introduced a coded elastic computing scheme by encoding both storage and download.
However, this method struggles with straggler mitigation and suffers from high encoding complexity and upload cost.
To address these issues, \cite{XJM2023} proposed an uncoded storage and Lagrange-coded download approach, requiring a large storage size. While the storage size is reduced in \cite{XJM2024}, further reductions are possible using coding techniques. 
Though it can be addressed by our work \cite{XSJM2025LCSD} which encodes both storage and download, the encoding complexity and decoding complexity increase. 
These limitations motivate us to develop a new coded elastic computing scheme that retains the low storage size achieved in \cite{XSJM2025LCSD} while reducing encoding complexity, computational complexity, upload cost, and decoding complexity, as demonstrated in \cite{XJM2023}.

In this paper, we introduce a new class of Lagrange-coded storage with uncoded download (LCSUD) schemes. Our key contributions are as follows.
\begin{enumerate}
    \item Unlike all existing Lagrange-coded elastic computing methods, this paper employs an uncoded download strategy while applying coding only to the storage.  This approach reduces storage size while maintaining low encoding and decoding complexity. 

    \item We propose three distinct LCSUD schemes that effectively address elasticity and straggler challenges. Among the three schemes, Scheme $1$ achieves the lowest download cost. Scheme $2$ achieves the lowest storage size and encoding complexity. Scheme $3$ has the lowest download cost, storage size and encoding complexity, while with higher upload cost and decoding complexity.

    \item Comparison between the proposed schemes and existing schemes shows that our schemes achieve the lowest storage size and upload cost, as shown in Section \ref{sec-discu}.  
\end{enumerate}

\paragraph*{Notation Convention}
$[N] = \{1, 2, \cdots, N\}$. 
We use $|\cdot|$ to represent the cardinality of a set.
$\mathbb{F}$ is a finite field.

\section{System Model}
The system consists of a master node and a set of $N$ worker machines, indexed by $[N]$. During the {\em storage placement phase}, each machine retains a processed version of the data matrix $\boldsymbol{A}  \in \mathbb{F}^{q\times v}$, with the storage size per machine normalized by the matrix size of $\boldsymbol{A}$. In the $t$-th time step, the input matrix is $\boldsymbol{B}^{(t)} \in \mathbb{F}^{ v \times r}$. The master assigns computation tasks to a set of available machines, denoted by $\mathcal{N}^{(t)}$, where $\mathcal{N}^{(t)} \subseteq [N]$ and $|\mathcal{N}^{(t)}| = N^{(t)}$. In the {\em download phase}, each machine downloads a function of $\boldsymbol{B}^{(t)}$ according to its task assignment. The download cost per machine during this phase is the size of the function it downloads. Following this, during the {\em computing phase}, each machine executes its assigned tasks locally and uploads the results back to the master node. The upload cost per machine corresponds to the size of the computation results it transmits. In the {\em decoding phase}, the master node collects sufficient results to reconstruct $\boldsymbol{AB}^{(t)}$, ensuring that the system can tolerate up to $S$ stragglers without delaying the process.
Let $L$ represent the recovery threshold, which is the smallest number of machines required for decoding successfully. This implies that, for any given time step $t$, the condition $N^{(t)} \geq L + S$ must hold. Additionally, we define $U$ as the maximum number of machines that can be preempted while still maintaining system tolerance, meaning that $N^{(t)} \geq N - U$ must be satisfied for every time step $t$.
\begin{definition}
\label{def-ars}
    {\bf (Availability Realization Set)} Given $U$, where $0 \leq  U \leq N-(L+S)$, the {\em availability realization set} of the system is denoted by $\boldsymbol{\mathcal{N}}_{U} = \{\mathcal{N} :  \mathcal{N} \subseteq [N],  N - U \leq  |\mathcal{N}| \leq N \}$,
    where $\mathcal{N}$ is referred to as an {\em availability realization}.  
\end{definition}
A system that tolerates up to $U$ preempted machines supports all availability realizations in $\boldsymbol{\mathcal{N}}_{U}$, i.e., $\mathcal{N}^{(t)} \in \boldsymbol{\mathcal{N}}_{U}$ for any time step $t$. Preempted machines are known before the download phase and are not assigned computation tasks, while stragglers are unknown in advance.

We begin by considering the case where the availability realization is fixed across all time steps. To illustrate our LCSUD schemes and explain their differences, we provide three examples. Following this, we introduce the general schemes. Next, we explore a scenario in which the system supports LCSUD schemes for any availability realization in $\boldsymbol{\mathcal{N}}_U$ for a given $U$.

\section{Three Examples with A Fixed $\mathcal{N}^{(t)}$}
\label{sec-examples}
Consider a system with $N = 6$, $L = 2$, $S = 1$, and $U = 0$. We have $\boldsymbol{\mathcal{N}}_{0} = \{ [6]\}$, i.e., for any time step $t$ we have $\mathcal{N}^{(t)} = [6]$. From $L = 2 $ and $N = 6$,  we consider $\beta_{1}$, $\beta_{2} \in \mathbb{F}$ and $\{\alpha_{n} \in \mathbb{F} : n \in [6]\}$, where $\{\alpha_{n} : n \in [6]\} \cap \{\beta_{1}, \beta_{2}\} = \emptyset$. Machine $n \in [6]$ corresponds to the number $\alpha_{n}$. 
We define $6$ sets of machines as the \emph{computation assignment}:
 \begin{equation}
  \label{eq-ex-assignment}
 \begin{split}
     \mathcal{W}_{1} = \{1, 2, 3\}, \mathcal{W}_{2} = \{2, 3, 4\}, \mathcal{W}_{3} = \{3, 4, 5\}, \\
     \mathcal{W}_{4} = \{4, 5, 6\}, \mathcal{W}_{5} = \{5, 6, 1\}, \mathcal{W}_{6} = \{6, 1, 2\}.
 \end{split}
 \end{equation}
Let  $\mathcal{L}_{g}$ be any subset of $\mathcal{W}_{g}$ with $|\mathcal{L}_{g}| = 2$.
The computation assignment in \eqref{eq-ex-assignment} will be used in the three examples.
In the following Example \ref{ex-low-commu}, it specifies the download of machines, where each machine receives half of matrix $\boldsymbol{B}^{(t)}$, with a download cost per machine of $\frac{vr}{2}$.
 
\subsection{Example for LCSUD Scheme $1$: Reduce Download Cost}
\begin{example}
\label{ex-low-commu}
The data matrix $\boldsymbol{A}$ is partitioned row-wise into two equal-sized sub-matrices, represented as $\boldsymbol{A} = [\boldsymbol{A}^T_1$, $\boldsymbol{A}^T_2]^T$.
We generate the  polynomial of degree $1$,
\begin{equation}
    \label{eq-ex1-encoding}
    X(z) = \boldsymbol{A}_{1} \cdot \frac{z-\beta_{2}}{\beta_{1}-\beta_{2}} + \boldsymbol{A}_{2} \cdot \frac{z-\beta_{1}}{\beta_{2}-\beta_{1}},
\end{equation}
where $X(\beta_{l}) = \boldsymbol{A}_{l}$ for $l \in [2]$.
Let machine $n \in [6]$ store $X(\alpha_{n}) = \Tilde{\boldsymbol{A}}_{n}$.
During the download phase, the matrix $\boldsymbol{B}^{(t)}$ is divided column-wise into $6$ sub-matrices of equal size, represented as
$\boldsymbol{B}^{(t)} = [\boldsymbol{B}^{(t)}_{1}$,  $\boldsymbol{B}^{(t)}_{2}$, $\boldsymbol{B}^{(t)}_{3}$,  $\boldsymbol{B}^{(t)}_{4}$, $\boldsymbol{B}^{(t)}_{5}$, $\boldsymbol{B}^{(t)}_{6}]$. 
Based on \eqref{eq-ex-assignment}, machine $n \in \mathcal{W}_{g}$ downloads $\boldsymbol{B}^{(t)}_{g}$ for all $g \in [6]$. 
Specifically, machine $1$ downloads $\boldsymbol{B}^{(t)}_{g}$ for $g \in \{1$, $5$, $6\}$. 
Machine $2$ downloads $\boldsymbol{B}^{(t)}_{g}$ for $g \in \{1$, $2$, $6\}$. 
Machine $3$  downloads $\boldsymbol{B}^{(t)}_{g}$ for $g \in \{1$, $2$, $3\}$. 
Machine $4$ downloads $\boldsymbol{B}^{(t)}_{g}$ for $g \in \{2$, $3$, $4\}$. 
Machine $5$ downloads $\boldsymbol{B}^{(t)}_{g}$ for $g \in \{3$, $4$, $5\}$, 
and machine $6$ downloads $\boldsymbol{B}^{(t)}_{g}$ for $g \in \{4$, $5$, $6\}$, which is shown in Fig. \ref{fig-cyclic}.
\begin{figure}[H]
\centerline{\includegraphics[scale= 0.23]{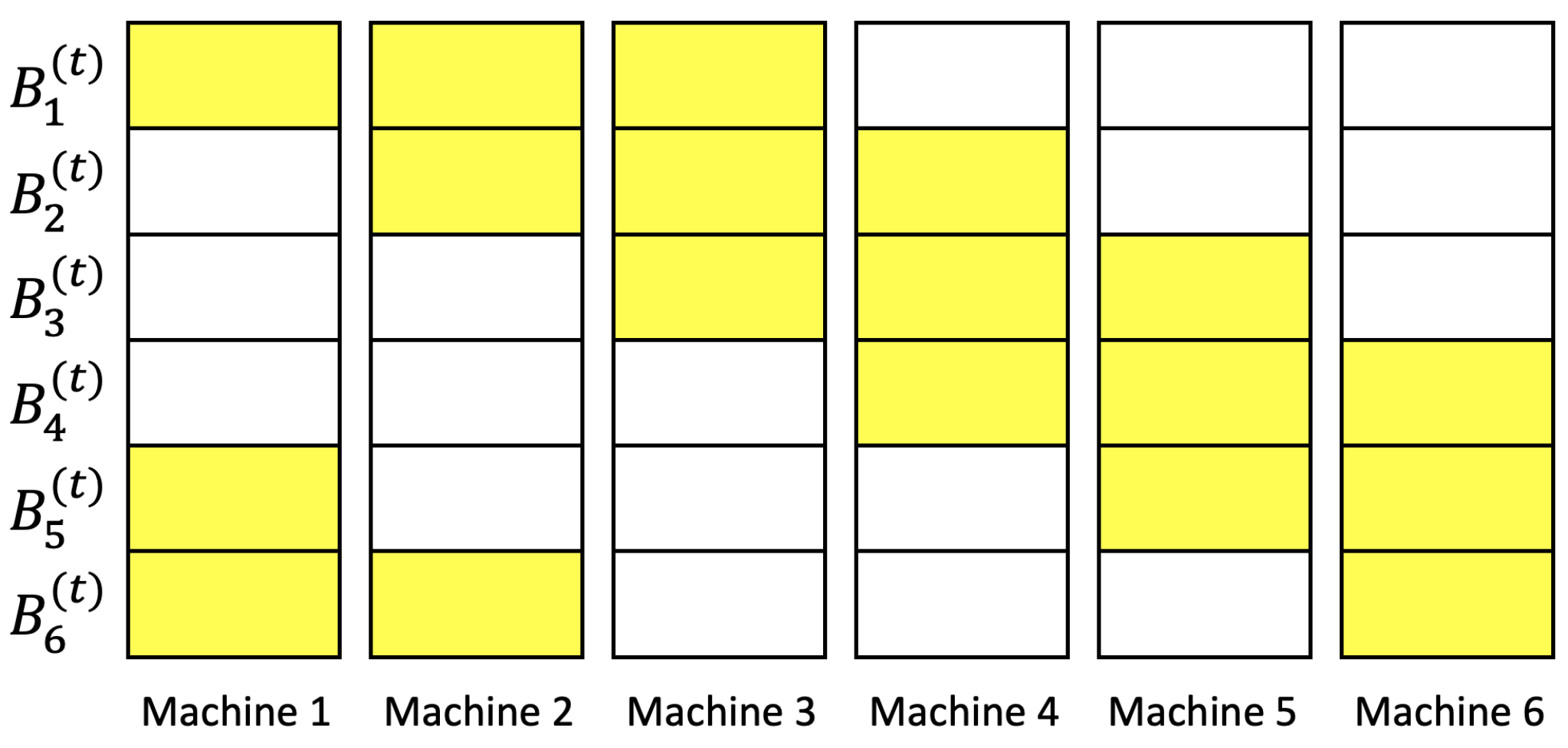}}
\caption{Downloads in Example \ref{ex-low-commu}, where the yellow shaded area represents the download data for the available machines.}
\label{fig-cyclic}
\end{figure}

In the computing phase, machine $n \in [6]$ computes 
$\{\Tilde{\boldsymbol{A}}_{n} \boldsymbol{B}^{(t)}_{g}:$ $n \in \mathcal{W}_{g}$, $g \in [6] \}$.
Recall that $ \boldsymbol{A} \boldsymbol{B}^{(t)} $ consists of $12$ sub-matrices, denoted by  $\{ \boldsymbol{A}_{l} \boldsymbol{B}^{(t)}_{g}:$ $l \in [2]$, $g \in [6]\}$.
The master will reconstruct $\{ \boldsymbol{A}_{l} \boldsymbol{B}^{(t)}_{g}:$ $l \in [2] \}$, by the computation results from machines $\mathcal{W}_{g}$, for each $g \in [6]$. Specifically, we define the following $6$ polynomials each of degree $1$, 
\begin{align}
    \label{eq-ex1-polys}
    F_{g}(z) = X(z) \cdot \boldsymbol{B}^{(t)}_{g}, \text{ for } g \in [6],
\end{align}
where $X(z)$ is defined in \eqref{eq-ex1-encoding}. 
For each polynomial $F_{g}(z)$, we have $\boldsymbol{A}_{l}\boldsymbol{B}^{(t)}_{g}$ $\overset{(a)}{=}$ $X(\beta_{l}) \boldsymbol{B}^{(t)}_{g}$ $\overset{(b)}{=}$ $F_{g}(\beta_{l})$ for $l \in [2]$, where $(a)$ is due to 
$\boldsymbol{A}_{l}  = X(\beta_{l})$ and $(b)$ is due to \eqref{eq-ex1-polys}. 
This means that $\boldsymbol{A}_{l}\boldsymbol{B}^{(t)}_{g}$ is an point of the polynomial $F_{g}(z)$. Also, the computation results uploaded by machines $\mathcal{W}_{g}$ are points on $F_{g}(z)$, as $\Tilde{\boldsymbol{A}}_{n} \boldsymbol{B}^{(t)}_{g}$ $\overset{(a)}{=}$ $X(\alpha_{n})\boldsymbol{B}^{(t)}_{g}$ $\overset{(b)}{=}$ $F_{g}(\alpha_{n})$ for $n \in \mathcal{W}_{g}$, where $(a)$ is due to  $\Tilde{\boldsymbol{A}}_{n}  = X(\alpha_{n})$,
and $(b)$ is due to \eqref{eq-ex1-polys}. 
Hence, recovering $\boldsymbol{A}_{l}\boldsymbol{B}^{(t)}_{g}$ for $l\in [2]$ and $g \in [6]$ equals to evaluating the unknown points $F_{g}(\beta_{l})$, using the known points, i.e., computation results.
Using Lagrange interpolation, the master computes 
$\boldsymbol{A}_{l}\boldsymbol{B}^{(t)}_{g}$ $=$ $F_{g}(\beta_{l})$ $=$ $\sum_{n \in \mathcal{L}_g}  \Tilde{\boldsymbol{A}}_{n} \boldsymbol{B}^{(t)}_{g}  \cdot$ $\prod_{n' \in \mathcal{W}_{g} \setminus \{n\}} \frac{\beta_{l}- \alpha_{n'}}{\alpha_{n}-\alpha_{n'}}$. 
By obtaining $\boldsymbol{A}_{l}\boldsymbol{B}^{(t)}_{g}$ for all $g \in [6]$ and $l\in[2]$, the master reconstructs  $\boldsymbol{A}\boldsymbol{B}^{(t)}$. 
Notably, $|\mathcal{L}_{g}| = L$ guarantees the successful decoding of $\boldsymbol{A}_{l}\boldsymbol{B}^{(t)}_{g}$, as the degree of $F_{g}(z)$ is $L-1$. The computation assignment $|\mathcal{W}_{g}| - |\mathcal{L}_{g}| = 1$ ensures the system can tolerate one straggler.
\end{example}

In this example, the storage size and download cost per machine are $\frac{1}{2}$ and $\frac{vr}{2}$, respectively, where each machine downloads $3$ out $6$ sub-matrices each of size $v \times \frac{r}{6}$. In the following Example \ref{ex-low-storage}, we consider the scenario where the system has smaller storage capacity and efficient transmission.
In this case, we can reduce the storage size per machine to $\frac{1}{4}$, while increasing the download cost per machine to $vr$.
\addtolength{\topmargin}{0.01in}

\subsection{Example for LCSUD Scheme $2$: Reduce Storage Size}
\begin{example}
\label{ex-low-storage}
In contrast to  Example \ref{ex-low-commu}, each machine $n \in [6]$ stores some sub-matrices of $\Tilde{\boldsymbol{A}}_{n}$. 
In particular, each $\boldsymbol{A}_{l}$ is further split row-wise into $6$ equal-sized sub-matrices for $l \in [2]$, denoted by 
$\boldsymbol{A}_{l} =$ $[\boldsymbol{A}^T_{l,1}$, $\boldsymbol{A}^T_{l,2}$, $\boldsymbol{A}^T_{l,3}$, $\boldsymbol{A}^T_{l,4}$, $\boldsymbol{A}^T_{l,5}$, $\boldsymbol{A}^T_{l,6}]^T$.
We consider the following $6$ polynomials, 
\begin{equation}
    \label{eq-ex2-encoding}
    X'_{g}(z) = \boldsymbol{A}_{1, g} \cdot \frac{z-\beta_{2}}{\beta_{1}-\beta_{2}} + \boldsymbol{A}_{2, g} \cdot \frac{z-\beta_{1}}{\beta_{2}-\beta_{1}}, \text{ for } g\in [6],
\end{equation}
where $X'_{g}(\beta_{l}) = \boldsymbol{A}_{l,g}$ for $l \in [2]$ and $g \in [6]$.
Based on \eqref{eq-ex-assignment}, machine $n \in [6]$ stores $\{ X'_{g}(\alpha_{n}): n \in \mathcal{W}_{g}, g \in [6]\}$. We denote $X'_{g}(\alpha_{n}) = \Tilde{\boldsymbol{A}}_{n,g}$.
The storage size per machine is $\frac{1}{4}$, as each machine stores $3$ coded matrices, each of size  $\frac{q}{12}\times v$. 

In the download phase, each machine downloads the entire $\boldsymbol{B}^{(t)}$.
In the computing phase, each machine $n \in [6]$ computes $\Tilde{\boldsymbol{A}}_{n,g} \boldsymbol{B}^{(t)}$ for $ n \in \mathcal{W}_{g}$ and $g \in [6]$. 

Recall that $\boldsymbol{A} \boldsymbol{B}^{(t)}$ contains $12$ sub-matrices, i.e.,  $\! \{\boldsymbol{A}_{l,g} \boldsymbol{B}^{(t)}\!:$ $l \in [2]$, $g \in [6]\}$. Next, using the computation results from machines $\mathcal{W}_{g}$, the master decodes $\{\boldsymbol{A}_{l,g} \boldsymbol{B}^{(t)}:$ $l \in [2] \}$ for each $g \in [6]$. We define the following polynomials,
\begin{align}
    \label{eq-ex2-polys}
    F'_{g}(z) = X'_{g}(z) \cdot \boldsymbol{B}^{(t)}, \text{ for } g \in [6],
\end{align}
where $X'_{g}(z)$ is defined in \eqref{eq-ex2-encoding}. 
For $g \in [6]$ and $l \in [2]$, $\boldsymbol{A}_{l,g} \boldsymbol{B}^{(t)}$ is the points of the polynomial $F'_{g}(z)$, as $\boldsymbol{A}_{l,g} \boldsymbol{B}^{(t)}$ $\overset{(a)}{=}$ $X'_{g}(\beta_{l}) \boldsymbol{B}^{(t)}$ $\overset{(b)}{=}$ $ F'_{g}(\beta_{l})$, where $(a)$ is due to 
$\boldsymbol{A}_{l,g}  = X'_{g}(\beta_{l})$, and $(b)$ is due to \eqref{eq-ex2-polys}. 
Also, the computation results uploaded by $\mathcal{W}_{g}$ are points on $F'_{g}(z)$, as $\Tilde{\boldsymbol{A}}_{n,g} \boldsymbol{B}^{(t)}$ $\overset{(a)}{=}$ $X'_{g}(\alpha_{n}) \boldsymbol{B}^{(t)}$ $\overset{(b)}{=}$ $F'_{g}(\alpha_{n})$ for $n$ $\in \mathcal{W}_{g}$, where $(a)$ is due to $\Tilde{\boldsymbol{A}}_{n,g} = X'_{g}(\alpha_{n})$, and $(b)$ is due to \eqref{eq-ex2-polys}. 
From Lagrange interpolation, the master evaluates $\boldsymbol{A}_{l,g} \boldsymbol{B}^{(t)}$ using the known points, i.e., computation results, by computing $\boldsymbol{A}_{l,g} \boldsymbol{B}^{(t)}$  $= F'_{g}(\beta_{l})$ $=$ $\sum_{n \in \mathcal{L}_g}  \Tilde{\boldsymbol{A}}_{n,g} \boldsymbol{B}^{(t)}$ $\cdot$ $\prod_{n' \in \mathcal{W}_{g} \setminus \{n\}} \frac{\beta_{l}- \alpha_{n'}}{\alpha_{n}-\alpha_{n'}}$.  
By obtaining $\boldsymbol{A}_{l,g}\boldsymbol{B}^{(t)}$ for $g \in [6]$ and $l\in[2]$, the master successfully reconstructs $\boldsymbol{A}\boldsymbol{B}^{(t)}$. 
\end{example}

In the following Example \ref{ex-low-both}, we consider a system that reduces both storage size and download cost. The storage size is maintained at $\frac{1}{4}$, while the download cost is reduced to $\frac{vr}{2}$.
However, the division strategies for $\boldsymbol{A}$ and $\boldsymbol{B}^{(t)}$ are modified, resulting in increased upload cost and decoding complexity due to the larger size of the computation results.

\subsection{Example for LCSUD Scheme $3$: Reduce both Storage Size and Download Cost with Higher Upload Cost}
\begin{example}
\label{ex-low-both}
In contrast to Example \ref{ex-low-storage}, each $\boldsymbol{A}_{l}$ for $l \in [2]$ is split column-wise into $6$ sub-matrices of equal size, denoted by $\boldsymbol{A}_{l} = $ $[ \boldsymbol{A}_{l,1}$,  $\boldsymbol{A}_{l,2}$, $\boldsymbol{A}_{l,3}$, $\boldsymbol{A}_{l,4}$, $\boldsymbol{A}_{l,5}$, $\boldsymbol{A}_{l,6}]$. 
Next, $\boldsymbol{B}^{(t)}$ is split row-wise into $6$ equal-sized sub-matrices, denoted by $\boldsymbol{B}^{(t)} = [(\boldsymbol{B}^{(t)}_{1})^T$, $(\boldsymbol{B}^{(t)}_{2})^T$, $(\boldsymbol{B}^{(t)}_{3})^T$, $(\boldsymbol{B}^{(t)}_{4})^T$, $(\boldsymbol{B}^{(t)}_{5})^T$, $(\boldsymbol{B}^{(t)}_{6})^T]^T$.
Consider the following $6$ polynomials,
\begin{equation}
    \label{eq-ex3-encoding}
    X''_{g}(z) = \boldsymbol{A}_{1, g} \cdot \frac{z-\beta_{2}}{\beta_{1}-\beta_{2}} + \boldsymbol{A}_{2, g} \cdot \frac{z-\beta_{1}}{\beta_{2}-\beta_{1}}, \text{ for } g\in [6]. 
\end{equation}
We have $X''_{g}(\beta_{l}) = \boldsymbol{A}_{l,g}$ for $l \in [2]$ and $g \in [6]$.
Based on \eqref{eq-ex-assignment}, each machine $n \in [6]$ stores $\{ X''_{g}(\alpha_{n}): n \in \mathcal{W}_{g}, g \in [6]\}$. We denote $X''_{g}(\alpha_{n}) = \Tilde{\boldsymbol{A}}_{n,g}$.
In the download phase, machine $n \in [6]$ downloads $\{ \boldsymbol{B}^{(t)}_{g}:$ $n \in \mathcal{W}_{g}$, $g \in [6]\}$.
In the computing phase, the computation tasks of machine $n$ are $\Tilde{\boldsymbol{A}}_{n,g} \boldsymbol{B}^{(t)}_{g}$ for $n \in \mathcal{W}_{g}$ and $g \in [6]$. 
Each computation result has the size of $\frac{q}{2} \times r$, which is larger than the $\frac{q}{12} \times r$ in Example \ref{ex-low-storage}, and the $\frac{q}{2} \times \frac{r}{6}$ in Example \ref{ex-low-commu}.
Recall that $\boldsymbol{AB}^{(t)}$ consists of  $12$ blocks, $\boldsymbol{A}_{l,g}\boldsymbol{B}^{(t)}_{g}$ for $l \in [2]$ and $g \in [6]$.
We define 
\begin{align}
    \label{eq-ex3-polys}
    F''_{g}(z) = X''_{g}(z) \cdot \boldsymbol{B}^{(t)}_{g}, \text{ for } g \in [6],
\end{align}
where $X''_{g}(z)$ is defined in \eqref{eq-ex3-encoding}. 
For $g \in [6]$ and $l \in [2]$, $\boldsymbol{A}_{l,g} \boldsymbol{B}^{(t)}_{g}$ is the points on $F''_{g}(z)$, as $\boldsymbol{A}_{l,g} \boldsymbol{B}^{(t)}_{g}$ $\overset{(a)}{=}$ $X''_{g}(\beta_{l})$ $\boldsymbol{B}^{(t)}_{g}$ $\overset{(b)}{=}$ $F''_{g}(\beta_{l})$, where $(a)$ is due to 
$\boldsymbol{A}_{l,g} = X''_{g}(\beta_{l})$,
and $(b)$ is due to \eqref{eq-ex3-polys}. Similarly, the computation results from  $\mathcal{W}_{g}$ are points on $F''_{g}(z)$, as $\Tilde{\boldsymbol{A}}_{n,g} \boldsymbol{B}^{(t)}$ $\overset{(a)}{=}$ $X''_{g}(\alpha_{n})$ $\boldsymbol{B}^{(t)}$ $\overset{(b)}{=}$ $F''_{g}(\alpha_{n})$ for $n \in \mathcal{W}_{g}$, where $(a)$ is due to 
$\Tilde{\boldsymbol{A}}_{n,g} = X''_{g}(\alpha_{n})$,
and $(b)$ is due to \eqref{eq-ex3-polys}.
Hence, using Lagrange interpolation, the master computes $\boldsymbol{A}_{l,g} \boldsymbol{B}^{(t)}_{g}$ $=$ $F''_{g}(\beta_{l})$ $=$ $\sum_{n \in \mathcal{L}_g}  \Tilde{\boldsymbol{A}}_{n,g} \boldsymbol{B}^{(t)}_{g}$ $\cdot$ $\prod_{n' \in \mathcal{W}_{g} \setminus \{n\}} \frac{\beta_{l}- \alpha_{n'}}{\alpha_{n}-\alpha_{n'}}$. 
By obtaining $\boldsymbol{A}_{l,g}\boldsymbol{B}^{(t)}_{g}$ for $g \in [6]$ and $l\in[2]$, the master successfully reconstructs $\boldsymbol{A}\boldsymbol{B}^{(t)}$. 
\end{example}

\subsection{Difference Among Three Schemes}
It can be seen that the three examples discussed above share the same computation assignment, $\mathcal{W}_{1}$, $\mathcal{W}_{2}, \cdots, \mathcal{W}_{6}$. In order to achieve the assigned computation tasks on each machine, the storage and download can be different. In detail, in Scheme $1$ shown in Example~\ref{ex-low-commu}, the downloads of machine are as shown Fig.~\ref{fig-cyclic} based on the computation assignment. In this case, each machine $n$ downloads $3$ sub-matrices of $\boldsymbol{B}^{(t)}$ (low download cost) while storing an entire coded matrix $\Tilde{\boldsymbol{A}}_{n}$ (high storage size). In Example \ref{ex-low-storage}, the storage placement is based on the computation assignment. In this case, each machine $n$ downloads the entire $\boldsymbol{B}^{(t)}$ (high download cost) while storing $3$ sub-matrices of $\Tilde{\boldsymbol{A}}_{n}$ (low storage size). In Example \ref{ex-low-both}, both storage placement and download  based on the computation assignment. In this case, each machine $n$ stores sub-matrices of $\Tilde{\boldsymbol{A}}_{n}$ (low storage size) and receives sub-matrices of $\boldsymbol{B}^{(t)}$ (low download cost). However, the upload cost and decoding complexity are increased.

\section{Proposed General LCSUD Systems}
\label{sec-fix}
We first consider a system with a fixed available realization, i.e., $\mathcal{N}^{(t)}$ remains consistent across all time steps. We then consider the system where $\mathcal{N}^{(t)}$ changes over time. 

\subsection{General LCSUD Schemes for A Fixed $\mathcal{N}^{(t)}$}
\label{sec: LCSUD Schemes 1}
Consider $L$ distinct numbers $\{\beta_{l} \in \mathbb{F} : l \in [L]\}$ and $N$  distinct numbers $\{\alpha_{n} \in \mathbb{F}:$  $n \in [N]\}$, where $\{\alpha_{n} :$ $n \in [N]\}$ $\cap$ $\{\beta_{l} :$ $l \in [L]\}$ $=$ $\emptyset$. 
Each machine $n \in$ $[N]$ corresponds to the number $\alpha_{n}$. 
We denote $i_{n}$ as the $n$-th machine in $\mathcal{N}^{(t)}$. 
We generate $N^{(t)}$ sets of machines, $\mathcal{W}^{(t)}_{1}$, $\mathcal{W}^{(t)}_{2}$, $\cdots$, $\mathcal{W}^{(t)}_{N^{(t)}}$, denoted as the computation assignment. Each set $\mathcal{W}_{g}^{(t)}$ for $g \in [N^{(t)}]$ is defined as $\mathcal{W}_{g}^{(t)}$ $=$ $\{ i_{g \% N^{(t)}}$,   $i_{(g+1) \% N^{(t)}}$, $\cdots$,  $i_{(g+L+S-1) \% N^{(t)}}\}$. Here, we define $a \% N^{(t)} = a - N^{(t)} \lfloor \frac{a-1}{N^{(t)}}\rfloor$.
We denote $\mathcal{L}_{g}$ as any subset of $\mathcal{W}_{g}$ with $|\mathcal{L}_{g}| = L$.

\subsubsection{LSCUD Scheme $1$} {\bf(Reduce Download Cost)} Data matrix $\boldsymbol{A}$ is split row-wise into $L$ equal-sized sub-matrices, denoted by 
$\boldsymbol{A} = [\boldsymbol{A}^T_{1}$, $\boldsymbol{A}^T_{2}$, $\cdots$, $\boldsymbol{A}^T_{L}]^T$. Consider
\begin{equation}
    \label{eq-encoding1}
    X(z) = \sum_{l \in [L]} \boldsymbol{A}_{l} \cdot \prod_{l'\in [L] \setminus {\{l\}}} {\frac{z-\beta_{l'}}{\beta_{l}-\beta_{l'}}},
\end{equation}
where $X(\beta_{l}) = \boldsymbol{A}_{l}$ for $l \in [L]$.
Each machine $n \in \mathcal{N}^{(t)}$ stores $X(\alpha_{n})$, where we define $X(\alpha_{n}) = \Tilde{\boldsymbol{A}}_{n}$.

In the download phase, matrix $\boldsymbol{B}^{(t)}$ is split column-wise into $N^{(t)}$  of equal-sized sub-matrices, denoted by 
$\boldsymbol{B}^{(t)}$ $=$ $[\boldsymbol{B}^{(t)}_1$, $\boldsymbol{B}^{(t)}_2$, $\cdots$, $\boldsymbol{B}^{(t)}_{N^{(t)}}]$.
Each machine $n \in \mathcal{N}^{(t)}$ downloads $\{\boldsymbol{B}^{(t)}_{g}:$ $n \in \mathcal{W}_{g}$, $g \in [N^{(t)}]\}$.
In the computing phase, machine $n \in \mathcal{N}^{(t)}$ computes $\{ \Tilde{\boldsymbol{A}}_{n} \boldsymbol{B}^{(t)}_{g} :$ $n \in \mathcal{W}_{g}$, $g \in [N^{(t)}]\}$.

In the decoding phase, we define the following polynomials,
\begin{align}
    \label{eq-1-polys}
    F_{g}(z) = X(z) \cdot \boldsymbol{B}^{(t)}_{g}, \text{ for } g \in [N^{(t)}],
\end{align}
where $X(z)$ is defined as \eqref{eq-encoding1}. 
We have $\boldsymbol{A}_{l}\boldsymbol{B}^{(t)}_{g}$ $\overset{(a)}{=}$ $X(\beta_{l}) \boldsymbol{B}^{(t)}_{g}$ $\overset{(b)}{=}$ $F_{g}(\beta_{l})$ for $l \in [L]$ and $g \in [N^{(t)}]$, where $(a)$ is due to $\boldsymbol{A}_{l} = X(\beta_{l})$,
and $(b)$ is due to \eqref{eq-1-polys}. Also, we have $\Tilde{\boldsymbol{A}}_{n}\boldsymbol{B}^{(t)}_{g}$ $\overset{(a)}{=}$ $X(\alpha_{n})\boldsymbol{B}^{(t)}_{g}$ $\overset{(b)}{=}$ $F_{g}(\alpha_{n})$ for $n\in \mathcal{W}_{g}$, where $(a)$ is due to 
$\Tilde{\boldsymbol{A}}_{n} = X(\alpha_{n})$,
and $(b)$ is due to \eqref{eq-1-polys}. 
Using Lagrange interpolation, the master computes $ \boldsymbol{A}_{l}\boldsymbol{B}^{(t)}_{g}$ $=$ $F_{g}(\beta_{l})$ $=$ $\sum_{n \in \mathcal{L}_g}$  $\Tilde{\boldsymbol{A}}_{n} \boldsymbol{B}^{(t)}_{g}$  $\cdot$  $\prod_{n' \in \mathcal{W}_{g} \setminus \{n\}} \frac{\beta_{l} - \alpha_{n'}}{\alpha_{n}-\alpha_{n'}}$.
By obtaining $\boldsymbol{A}_{l}\boldsymbol{B}^{(t)}_{g}$ for all $l 
\in [L]$ and $g \in [N^{(t)}]$, the master  successfully reconstructs $\boldsymbol{AB}^{(t)}$.

\subsubsection{Scheme $2$} {\bf (Reduce Storage Size)} The data matrix $\boldsymbol{A}$ is split row-wise into $L$ of equal-sized sub-matrices, denoted by $\boldsymbol{A} = [\boldsymbol{A}^T_{1}$, $\boldsymbol{A}^T_{2}$, $\cdots$, $\boldsymbol{A}^T_{L}]^T$. Additionally, each $\boldsymbol{A}_{l}$ is further divided row-wise into $N^{(t)}$ sub-matrices of equal size,  i.e., $\boldsymbol{A}_{l}$ $=$ $[\boldsymbol{A}^T_{l,1}$, $\boldsymbol{A}^T_{l,2}$, $\cdots$, $\boldsymbol{A}^T_{l,N^{(t)}}]^T$. 
We consider
\begin{equation}
    \label{eq-2-encoding}
    X'_{g}(z) = \sum_{l \in [L]} \boldsymbol{A}_{l,g} \cdot \prod_{l'\in [L] \setminus {\{l\}}} {\frac{z-\beta_{l'}}{\beta_{l}-\beta_{l'}}}, \text{ for } g\in [N^{(t)}], 
\end{equation}
which satisfies $X'_{g}(\beta_{l}) = \boldsymbol{A}_{l, g}$, for $l \in [L]$.
Each machine $n \in \mathcal{N}^{(t)}$ stores $\{X'_{g}(\alpha_{n}) :$ $n \in \mathcal{W}_{g}$, $g \in [N^{(t)}]\}$, where we define $X'_{g}(\alpha_{n}) = \Tilde{\boldsymbol{A}}_{n, g}$.

In the download phase, each available machine downloads the entire $\boldsymbol{B}^{(t)}$.
In the computing phase, machine $n \in \mathcal{N}^{(t)}$ computes $\{\Tilde{\boldsymbol{A}}_{n,g} \boldsymbol{B}^{(t)} : n \in \mathcal{W}_{g}, g \in [N^{(t)}] \}$.

In the decoding phase, we define the following polynomials,
\begin{align}
    \label{eq-2-polys}
    F'_{g}(z) = X'_{g}(z) \cdot \boldsymbol{B}^{(t)}, \text{ for } g \in [N^{(t)}],
\end{align}
where $X(z)$ is defined in \eqref{eq-2-encoding}.
We have $\boldsymbol{A}_{l, g}\boldsymbol{B}^{(t)}$ $\overset{(a)}{=}$ $X'_{g}(\beta_{l}) \boldsymbol{B}^{(t)}$ $\overset{(b)}{=}$ $F'_{g}(\beta_{l})$ for $l \in [L]$ and $g \in [N^{(t)}]$, where $(a)$ is due to 
$\boldsymbol{A}_{l, g} = X'_{g}(\beta_{l})$,
and $(b)$ is due to \eqref{eq-2-polys}. 
Also, we have $\Tilde{\boldsymbol{A}}_{n, g} \boldsymbol{B}^{(t)}$ $\overset{(a)}{=}$ $X'(\alpha_{n})\boldsymbol{B}^{(t)}$ $\overset{(b)}{=}$ $F'_{g}(\alpha_{n})$ for $n\in \mathcal{W}_{g}$, where $(a)$ is due to 
$\Tilde{\boldsymbol{A}}_{n, g} = X'(\alpha_{n})$,
and $(b)$ is due to \eqref{eq-2-polys}. 
Using Lagrange interpolation, the master computes
$\boldsymbol{A}_{l,g}\boldsymbol{B}^{(t)}$ $=$  $F'_{g}(\beta_{l})$ $=$ $\sum_{n \in \mathcal{L}_g}$  $\Tilde{\boldsymbol{A}}_{n,g} \boldsymbol{B}^{(t)}$ $\cdot$  $\prod_{n' \in \mathcal{W}_{g} \setminus \{n\}} \frac{\beta_{l} - \alpha_{n'}}{\alpha_{n}-\alpha_{n'}}$.
By obtaining $ \boldsymbol{A}_{l,g}\boldsymbol{B}^{(t)}$ for all 
$l \in [L]$ and $g \in [N^{(t)}]$, the master  successfully  reconstructs $\boldsymbol{AB}^{(t)}$.

\subsubsection{LCSUD Scheme $3$}
{\bf (Reduce Storage Size and Download Cost with Higher Upload Cost)}
The data matrix $\boldsymbol{A}$ is split row-wise into $L$ equal-sized sub-matrices, i.e., $\boldsymbol{A}$ $=$ $[\boldsymbol{A}^T_{1}$, $\boldsymbol{A}^T_{2}$, $\cdots$, $\boldsymbol{A}^T_{L}]^T$. Moreover, each $\boldsymbol{A}_{l}$ is further divided into $N^{(t)}$ sub-matrices column-wise,  denoted by $\boldsymbol{A}_{l}$ $=$ $[\boldsymbol{A}_{l,1}$, $\boldsymbol{A}_{l,2}$, $\cdots$, $\boldsymbol{A}_{l,N^{(t)}}]$. 
We generate
\begin{equation}
    \label{eq-3-encoding}
    X''_{g}(z) = \sum_{l \in [L]} \boldsymbol{A}_{l,g} \cdot \prod_{l'\in [L] \setminus {\{l\}}} {\frac{z-\beta_{l'}}{\beta_{l}-\beta_{l'}}}, \text{ for } g\in [N^{(t)}], 
\end{equation}
which satisfies $X''_{g}(\beta_{l}) = \boldsymbol{A}_{l, g}$, for $l \in [L]$.
Each machine $n \in \mathcal{N}^{(t)}$ stores $\{X''_{g}(\alpha_{n}) :$ $n \in \mathcal{W}_{g}, g \in [N^{(t)}]\}$, where we define $X''_{g}(\alpha_{n}) = \Tilde{\boldsymbol{A}}_{n, g}$.

In the download phase,  matrix $\boldsymbol{B}^{(t)}$ is split row-wise into $N^{(t)}$  equal-sized sub-matrices, denoted by $ \boldsymbol{B}^{(t)}$ $ =$ $[(\boldsymbol{B}^{(t)}_1)^T$, $(\boldsymbol{B}^{(t)}_2)^T$, $\cdots$, $(\boldsymbol{B}^{(t)}_{N^{(t)}})^T]^T$.
Each  machine $n \in \mathcal{N}^{(t)}$ downloads $\{\boldsymbol{B}^{(t)}_{g}:$ $n \in \mathcal{W}_{g}$, $g \in [N^{(t)}]\}$.
In the computing phase, machine $n \in \mathcal{N}^{(t)}$ computes $\{ \Tilde{\boldsymbol{A}}_{n,g} \boldsymbol{B}^{(t)}_{g}:$ $n \in \mathcal{W}_{g}, g \in [N^{(t)}]\}$.
Recall that $\boldsymbol{AB}^{(t)} $ contains $LN^{(t)}$ sub-matrices $\boldsymbol{A}_{l, g}\boldsymbol{B}^{(t)}$ for $l\in [L]$ and $g \in [N^{(t)}]$, each having dimensions $\frac{q}{L} \times r$.
In the decoding phase, we define 
\begin{align}
    \label{eq-3-polys}
    F''_{g}(z) = X''(z) \cdot \boldsymbol{B}^{(t)}_{g}, \text{ for } g \in [N^{(t)}],
\end{align}
where $X''(z)$ is defined in  \eqref{eq-3-encoding}. 
We have $\boldsymbol{A}_{l,g}\boldsymbol{B}^{(t)}_{g}$ $\overset{(a)}{=}$ $X''(\beta_{l}) \boldsymbol{B}^{(t)}_{g}$ $\overset{(b)}{=}$ $F''_{g}(\beta_{l})$ for $l \in [L]$ and $g \in [N^{(t)}]$, where $(a)$ is due to
$\boldsymbol{A}_{l,g} = X''(\beta_{l}) $,
and $(b)$ is due
to \eqref{eq-3-polys}.
Also,  we have $\Tilde{\boldsymbol{A}}_{n,g}\boldsymbol{B}^{(t)}_{g} \overset{(a)}{=} X''(\alpha_{n})\boldsymbol{B}^{(t)}_{g} \overset{(b)}{=} F''_{g}(\alpha_{n})$ for $n \in \mathcal{W}_{g}$, where $(a)$ is due to 
$\Tilde{\boldsymbol{A}}_{n,g} = X''(\alpha_{n}) $,
and $(b)$ is due to \eqref{eq-3-polys}. 
Using Lagrange interpolation, the master computes
$\boldsymbol{A}_{l,g}\boldsymbol{B}^{(t)}_{g}$ $=$ $F''_{g}(\beta_{l})$ $=$ $\sum_{n \in \mathcal{L}_g}  \Tilde{\boldsymbol{A}}_{n,g} \boldsymbol{B}^{(t)}_{g}$ $\cdot$ $\prod_{n' \in \mathcal{W}_{g} \setminus \{n\}}  \frac{\beta_{l}- \alpha_{n'}}{\alpha_{n}-\alpha_{n'}}$.
By obtaining $\boldsymbol{A}_{l,g}\boldsymbol{B}^{(t)}_{g}$ for $l \in [L]$ and $g \in [N^{(t)}]$, the master successfully reconstructs $\boldsymbol{AB}^{(t)}$.

Using Scheme $i$, where $i \in \{1,2,3\}$, the system can tolerate up to $S$ stragglers. Since $|\mathcal{W}_{g}| = L+S$, $L+S$ machines are assigned to decode some blocks. However,  $L$ machines are necessary for successful decoding, as the degrees of the polynomials $F(z)$, $F'(z)$, and $F''(z)$ are $L-1$.

\subsection{General LCSUD Schemes 
Given $\boldsymbol{\mathcal{N}}_{U}$}  
Given $U$ and the availability realization set $\boldsymbol{\mathcal{N}}_{U}$, the storage placement of a machine is defined as the union of its storage placements across all availability realizations in $\boldsymbol{\mathcal{N}}_{U}$.

\section{Discussion}
\label{sec-discu}
\begin{table*}
\caption{Computational Complexity}
\centering
\label{table-complexity}
\begin{tabular}{|c|c|c|c|c|c|c|c|c|} 
\hline
    & Storage Size & $\mathcal{C}_{\text{Encoding}}$  & $\mathcal{C}_{\text{Download}}$ & $\mathcal{C}_{\text{Computing}}$ & $\mathcal{C}_{\text{Upload}}$ & $\mathcal{C}_{\text{Decoding}}$\\ 
\hline
Scheme $1$  & $\frac{1}{L}$ & $qv$  & $\frac{vr(L+S)}{N^{(t)}}$ & {\color{red}$\frac{qvr(L+S)}{LN^{(t)}}$} & ${\color{red}\frac{qr(L+S)}{LN^{(t)}}}$ & $qrL$ \\
\hline
Scheme $2$ & ${\color{red}\frac{L+S}{LN^{(t)}}}$ & ${\color{red}\frac{qv(L+S)}{N^{(t)}}}$ & $vr$ & {\color{red}$\frac{qvr(L+S)}{LN^{(t)}}$} & ${\color{red}\frac{qr(L+S)}{LN^{(t)}}}$  & $qrL$ \\
\hline
Scheme $3$ & ${\color{red}\frac{L+S}{LN^{(t)}}}$ & ${\color{red}\frac{qv(L+S)}{N^{(t)}}}$ & $\frac{vr(L+S)}{N^{(t)}}$ & {\color{red}$\frac{qvr(L+S)}{LN^{(t)}}$} & $\frac{qr(L+S)}{L}$ & $qrLN^{(t)}$ \\
\hline
\hline
\cite{yang2018coded}  & $\frac{1}{L}$ & $qv$ &  $vr$ & {\color{red}$\frac{qvr(L+S)}{LN^{(t)}}$} & ${\color{red}\frac{qr(L+S)}{LN^{(t)}}}$  & $qrL$\\
\hline  
\cite{XJM2023}   & $1$  & {\color{red}$\frac{vr(L+S)}{N^{(t)}}$} & $\frac{vr(L+S)}{LN^{(t)}}$ & {\color{red}$\frac{qvr(L+S)}{LN^{(t)}}$} &  {\color{red}$\frac{qr(L+S)}{LN^{(t)}}$} &  $qrL$\\
\hline
 \cite{XJM2024}   & $\frac{L+S}{N^{(t)}}$  & $vr$ & $\frac{vr}{L}$ & {\color{red}$\frac{qvr(L+S)}{LN^{(t)}}$} &  {\color{red}$\frac{qr(L+S)}{LN^{(t)}}$} &  $qrL$\\
\hline                                                                                                               
\cite{yangCEC}  & $\frac{1}{L}$ & $qv+\frac{vrL}{N^{(t)}}$ &  ${\color{red}\frac{vr}{N^{(t)}}}$ & {\color{red}$\frac{qvr}{N^{(t)}}$} & $qrL$ & ${\color{red}\mathcal{O}(1)}$\\
\hline
\end{tabular}
\end{table*}

\vspace{-3mm}

\subsection{Storage Size}
\label{sec-discuss-size}
\begin{figure}[H]
\centerline{\includegraphics[scale= 0.45]{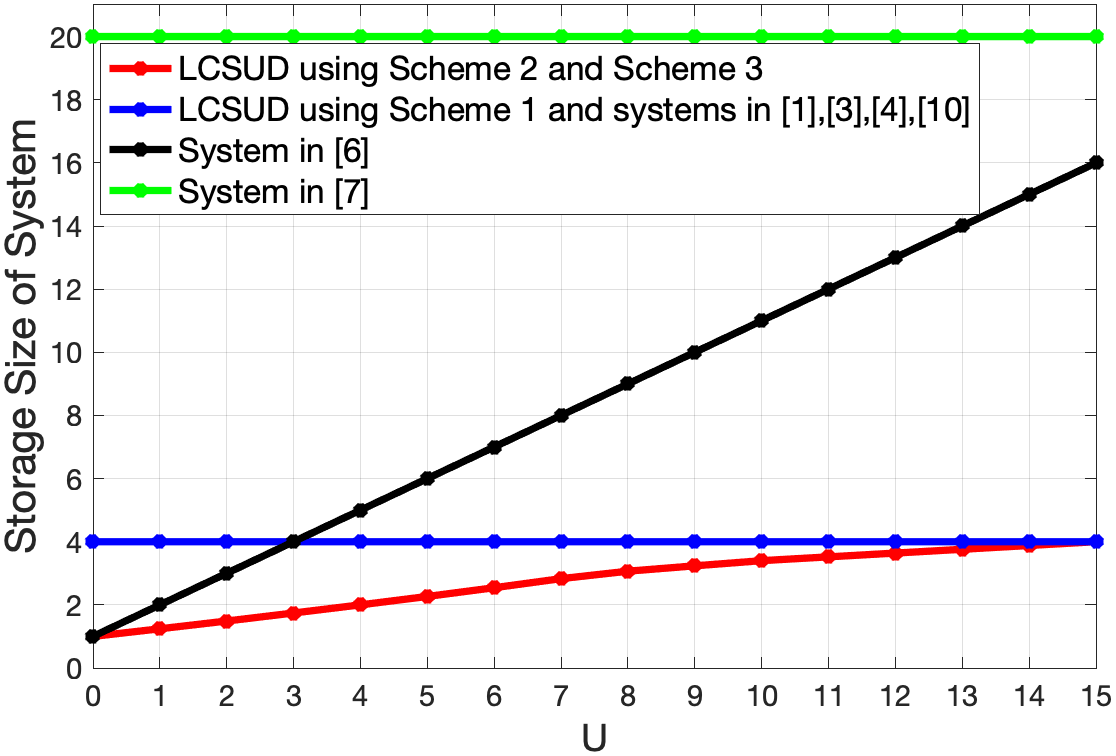}}
\caption{Storage Size of System when $N =20$, $L = 5$, and $S = 0$. The $x$-axis and $y$-axis represent $U$ and the storage size of the system normalized by the size of $\boldsymbol{A}$, respectively. } 
\label{fig-storage-saving}
\end{figure}

Using Scheme $2$ and Scheme $3$, the storage size per machine for a given $\mathcal{N}^{(t)}$ is $\frac{1}{L} \times \frac{L+S}{N^{(t)}} = \frac{L+S}{LN^{(t)}}$. For a given $U$, the storage size per machine increases to a value no greater than $\frac{1}{L}$. 
As a result, the overall storage size of the system does not exceed $\frac{N}{L}$. 
We now compare the storage size of the system using LCSUD with that of existing schemes.
Considering $N = 20$, $L= 5$, $S = 0$, and varying $U \in \{0, 1 , \cdots, 15\}$, the storage sizes of the system are illustrated in Fig. \ref{fig-storage-saving}.
\setlength\belowcaptionskip{-2ex}

In Fig. \ref{fig-storage-saving}, the blue line illustrates the storage size of the LCSUD system using Scheme $1$, as well as the systems in \cite{yang2018coded}, \cite{wcj2021hs}, \cite{myjpractice}, and \cite{yangCEC}, where each machine stores an entire coded matrix, resulting in a storage size of $\frac{N}{L} = 4$.
The black line corresponds to the minimum storage size of the system in \cite{usutec2022}, where at least $1+U$ machines store each row of the data matrix $\boldsymbol{A}$. 
The green line represents the scheme in \cite{XJM2023}, where each machine stores the entire data matrix $\boldsymbol{A}$, leading to a storage size of $1 \cdot N = 20$. The red line represents the LCSUD systems using Scheme $2$ and Scheme $3$, which achieve the lowest storage size compared to existing methods.
Interestingly, when $U = 15$, the storage size of Scheme $2$ and Scheme $3$ equals to the value represented by the blue line. This is because, in this case, the number of available machines may equal to the recovery threshold $L = 5$, meaning each machine $n$ must store the entire coded matrix $\Tilde{\boldsymbol{A}}_{n}$. 

\subsection{Computational Complexity} 
We compare the LCSUD schemes with several existing schemes designed for homogeneous systems. We denote encoding complexity at the master for each machine as $\mathcal{C}_\text{Encoding}$, download cost per machine as $\mathcal{C}_{\text{Download}}$, computing complexity per machine as $\mathcal{C}_{\text{Computing}}$, upload cost per machine as $\mathcal{C}_{\text{Upload}}$, and the decoding complexity at the master  as $\mathcal{C}_{\text{Decoding}}$. 
The comparisons are summarized in Table \ref{table-complexity}. The scheme proposed in \cite{yang2018coded}, although originally designed for matrix-vector multiplications, can be applied to matrix-matrix multiplications. The scheme in \cite{yangCEC}, listed in Table \ref{table-complexity}, is specifically designed for matrix-matrix multiplications. However, since it cannot tolerate stragglers, the computational complexity in \cite{yangCEC} is independent of $S$. Additionally, the best performance for each metric is highlighted in red, assuming $q = v = r$ and $S = 0$.
From Table \ref{table-complexity} and the discussion of storage size in Section \ref{sec-discuss-size}, we can draw the following conclusions. 

1) Compared to \cite{yang2018coded}, our Scheme $1$ achieves a lower download cost.
Scheme $2$ has the lower storage size and encoding complexity. 
Scheme $3$ has the lower storage size, encoding complexity and download cost, with higher upload cost and decoding complexity. 
2) Compared to \cite{yangCEC},  with $S = 0$  
Scheme $1$ has the lower encoding complexity and upload cost.
Scheme $2$ and $3$ have lower storage size, encoding complexity and upload cost.
3) Compared to \cite{XJM2023}, our schemes significantly reduce the storage size.
4) Compared to \cite{XJM2024}, Scheme $2$ and $3$ reduce the encoding complexity, with the smaller storage size.

\bibliographystyle{IEEEtran}
\bibliography{reference}
\end{document}